\documentstyle[aps,prl,epsf,twocolumn]{revtex}
\def\be{\begin{equation}}
\def\ee{\end{equation}}
\def\e{{\rm e}}
\begin{document}  
\draft

\title{Short time evolved wave functions for solving 
       quantum many-body problems}
\author{Orion Ciftja}
\address{Department of Physics, Prairie View A$\&$M University,
         Prairie View, Texas 77446 }
\author{Siu A. Chin}
\address{Department of Physics, Texas A$\&$M University, College
        Station, Texas 77843 }
\date{April 28, 2003}   
\maketitle

\begin{abstract}
The exact ground state of a strongly interacting quantum many-body system
can be obtained by evolving a trial state with finite overlap with
the ground state to infinite imaginary time. 
In many cases, since the convergence is exponential, the system converges 
essentially to the exact ground state in a relatively short time. 
Thus a short time evolved wave function 
can be an excellent approximation to the exact ground state. Such a 
short time evolved wave function can be obtained by factorizing, or 
splitting, the evolution operator to high order. 
However, for the imaginary time
Schr\"odinger equation, which contains an irreversible diffusion kernel,
all coefficients, or time steps, must be positive. (Negative time steps
would require evolving the diffusion process backward in time, which is 
impossible.) Heretofore, only second order factorization schemes can have
all positive coefficients, but without further iterations, 
these cannot be used to evolve the system long enough to be close to 
the exact ground state. 
In this work, we use a newly discovered fourth order positive factorization 
scheme which requires knowing both the potential and its gradient. 
We show that the resulting fourth order wave function alone, without further
iterations, gives an excellent description of strongly interacting quantum 
systems such as liquid $^4$He, comparable to the best variational results 
in the literature. This suggests that such a fourth order wave function can 
be used to study the ground state of diverse quantum many-body systems, 
including Bose-Einstein condensates and Fermi systems. 
\end{abstract}

\narrowtext

We consider a quantum system of $N$ particles with mass $m$ described by the
Hamiltonian:
\begin{equation}
{H}={T}+{V} \ \ ; \ \
{T}=-\lambda \sum_{i=1}^{N} \nabla_{i}^{2} \ \ ; \ \
{V}= \sum_{i>j}^{N} v(r_{ij})  \ ,
\label{eqH}
\end{equation}
where ${T}$ is the kinetic energy operator, ${V}$ is a 
sum of pair-wise potentials $v(r_{ij})$
and $\lambda=\hbar^2/(2 m)$.

In imaginary time $\tau=i \frac{t}{\hbar}$ the many-body 
time-dependent Schr\"{o}dinger equation can be written as
\begin{equation}
 -\frac{\partial}{\partial \tau} |\Psi(\tau) \rangle={H}
                                    |\Psi(\tau) \rangle \ ,
\label{imaginary}
\end{equation}
with formal solution
\begin{equation}
|\Psi(\tau) \rangle=\e^{-\tau {H}} |\Phi \rangle \ \ \ ; \ \ \
|\Phi \rangle \equiv |\Psi(0) \rangle  \ .
\label{formal}
\end{equation}
In the coordinate representation 
$ \Psi(R,\tau)=\langle R|\Psi(\tau) \rangle=
\langle R | \e^{-\tau {H}} |\Phi \rangle$, where
$R = \{ {\bf r}_1 \ldots {\bf r}_N \}$ denotes the set of 
all particle coordinates.
%
%
If the initial wave function $|\Phi \rangle$ is expanded
in the set of exact eigenfunctions
       $\left\{ \Phi_n \right\}$
of the Hamiltonian $ H$, then Eq.(\ref{formal}) has the more explicit form,
\begin{equation}
\Psi(R,\tau) 
= 
  \e^{-\tau \ E_0} \ \left[ c_0 \ \Phi_{0}+    
\sum_{n \neq 0}^{+\infty} c_n \ \e^{-\tau (E_n-E_0)} \ \Phi_n    \right]. 
\label{expand}
\end{equation}
Assuming the non-degeneracy of the ground state ($E_n-E_0 > 0 \ ; n \neq 0$),
the above wave function becomes proportional
to the exact ground state wave function in the limit of 
infinite imaginary time.
A basic strategy is then to start with a good trial wave function and
evolve it in imaginary time long enough to damp out all but
the exact ground state wave function. 

%
%
Since the imaginary time evolution cannot be done exactly, one
usually develops a short-time propagator by decomposing  
$\e^{-\tau {H}}=\e^{-\tau({T}+{V})}$ into exactly
solvable parts, and further iterate this short time propagator
to longer time. This is essentially the approach of the Diffusion
Monte Carlo (DMC) method\cite{rey,mosk,chindmc}. The need for iterations 
introduces the complication of branching, which is the hallmark of
Diffusion and Green's Function Monte Carlo methods\cite{gfmc}. 
Our idea is to develop a short time propagator via higher order 
decomposition that can be applied for a sufficient long time to project 
out an excellent approximation to the ground state {\it without} iteration.

First and second order factorization
schemes such as 
\begin{equation}
\e^{-\tau ({T}+{V})} \approx 
\left\{ 
\begin{array}{l}
\e^{-\tau {T}} \ \e^{-\tau {V}} +O(\tau^2)  \\
\e^{-\frac{1}{2}\tau {V}} \  \e^{-\tau {T}} \
\e^{-\frac{1}{2}\tau {V}} +O(\tau^3)  \ ,
\end{array}
\right.
\label{firstsecond}
\end{equation}
are well known, but without iterations, they cannot be applied at 
sufficiently large value of $\tau$ to get near to the ground state. 
It is also well known that in the context of 
symplectic integrators, the short time evolution operator can be 
factorized to arbitrarily high order in the 
form\cite{forest,creutz,candy,suzuki,yoshida,manda,mclachlan,koseleff}
\be
{\rm e}^{-\tau (T+V )}=\prod_i
{\rm e}^{-a_i\tau T}{\rm e}^{-b_i\tau V},
\label{arb}
\ee
with coefficients $\{a_i, b_i\}$ determined
by the required order of accuracy. However, as
first proved by Sheng\cite{sheng} and later by Suzuki\cite{suzukinogo} 
(using a more geometric argument), beyond second order, 
any factorization of the form 
(\ref{arb}) {\it must} contain some negative coefficients in the set 
$\{a_i, b_i\}$. Goldman and Kaper\cite{goldman}
later proved that any factorization of the form (\ref{arb}) must 
contain at least one negative coefficient for {\it both} operators. 
This means that for 
decompositions of the form (\ref{arb}), one must evolve the system backward in
time for some intermediate time steps. 
This is of little consequence for classical dynamics, or real time quantum 
dynamics, both of which are time-reversible. For
the imaginary time Schr\"odinger equation, whose kinetic energy operator 
is the time-irreversible diffusion kernel, this is detrimental. 
This is because 
${\rm e}^{-a_i\tau T}
\propto {\rm e}^{-({\bf r}^\prime-{\bf r})^2/(2a_i\tau)}$
is the diffusion Green's function. For positve $a_i$, this kernel can be 
simulated by Gaussian random walks. If $a_i$ were negative, 
the kernel would be unbound and unnormalizable, with no probabilistic based
(Monte Carlo) simulations possible. This is just a mathematical restatement
of the physical fact that diffusion is an irreversible process.
Positive decomposition coefficients are therefore absolutely 
essential for solving any evolution equation having an irreversible 
component, such as the imaginary time Schr\"odinger equation.
 
Since both classical and quantum dynamics are time-reversible, there is 
a lack of impetus to search for higher order factorization schemes with
only positive coefficients. While higher order factorizations of 
the form (\ref{arb}), with negative coefficients, 
have been studied extensively in the literature\cite{manda,mclachlan,koseleff},
it was only recently that Suzuki\cite{suzfour} 
and Chin\cite{chin} have found some fourth order (but no higher order) all 
forward time step decomposition schemes. In order to bypass Sheng and 
Suzuki's proof, one must introduce a higher order commutator $[V,[T,V]$
in additional to operators $T$ and $V$ used in (\ref{arb}). 
In this work, we use the fourth order factorization scheme\cite{suzfour,chin}
referred to as scheme A:

\begin{equation}
\e^{-\tau ({T}+{V})} =
\e^{ -\frac{1}{6}\tau V} \e^{-\frac{1}{2}\tau {T}} \
\e^{-\frac{2}{3}\tau \widetilde{V}} \
\e^{-\frac{1}{2}\tau {T}} \e^{-\frac{1}{6}\tau {V}}+O(\tau^5) \ ,
\label{china}
\end{equation}
with $\widetilde{V}$ given by
\begin{equation}
\widetilde{V}={V}+\frac{\tau^2}{48} 
                 \left[{V},[{T},{V}]\right]=
{V}+\frac{\tau^2}{48} 2 \lambda \sum_{i=1}^{N} |{\nabla}_i V |^2 \ .
\label{Vtilde}
\end{equation}
This scheme was also found by Koseloff\cite{kos}, but his coefficient for the
double commutator term is {\it incorrect} by a factor of three too large. 
For a more detailed discussion of positive factorization schemes and forward
symplector integrators, see Ref.\cite{chinsym}. 

To go from state vectors to coordinate wave functions, we 
insert complete sets of coordinate states, 
$1=\int dS \ |S \rangle \langle S |$ 
where $S=\{ {\bf s}_1 \ldots {\bf s}_N \}$
and write, for example, the operator equation (\ref{formal}) in the form
\begin{equation}
\Psi(R,\tau)=\int dS \  G(R,S,\tau) \ \Phi(S)  \ ,
\label{psigf}
\end{equation}
where the Green's function $G(R,S,\tau)$ is given by
\begin{equation}
G(R,S,\tau)= \langle R | \e^{-\tau {H}} | S \rangle.
\label{gf}
\end{equation}
The intermediate coordinates $S$ are also sometime referred
to as ``shadow" positions.
Each decomposition scheme then corresponds to a specific
wave function for the ground state.
For instance, the first-order scheme gives the {\bf linear}
wave function,
\begin{equation}
\Psi(R,\tau)=
\int dS \  \e^{-C \ (R-S)^2}  \ \e^{-\tau V(S)} \  \Phi(S)  \ ,
\label{psilinear}
\end{equation}
where  
$(R-S)^2\equiv\sum_{i=1}^{N} ({\bf r}_i-{\bf s}_i)^2$,
and where we have used the fundamental result that
the kinetic evolution operator is just the  
the diffusion Green's function, 
\begin{equation}
\langle R | \e^{-\tau {T}} | S \rangle \propto
\e^{-C \ (R-S)^2}  
\ \ ; \ \
C=\frac{1}{4 \tau \lambda}   \ .
\label{gaussian}
\end{equation}
Similarly, the second order scheme gives the following 
{\bf quadratic} wave function 
\begin{equation}
\Psi(R,\tau)=\e^{-\frac{\tau}{2} V(R)} \
\int dS \  \e^{-C \ (R-S)^2}  \
\e^{-\frac{\tau}{2} V(S)} \  \Phi(S)  \ .
\label{psiquadratic}
\end{equation}
%
%
Finally, the fourth order scheme A produces the following
{\bf quartic } many-body wave function,

\begin{eqnarray}
\Psi(R,\tau) &=&
\e^{-\frac{\tau}{6} V(R)}
\int dS^{\prime} \e^{-C^{\prime} (R-S^{\prime})^2} 
\e^{-\frac{2 \tau}{3} \widetilde V(S^{\prime})} \times    \nonumber \\
& &
\int dS \e^{-C^\prime (S^{\prime}-S)^2}
\e^{-\frac{\tau}{6} V(S)} \Phi(S)  \ ,
\label{psiquartic}
\end{eqnarray}
now with $ C^{\prime}=1/(2 \tau \lambda) $.
%
%

In all these wave functions, there is only a single paramater, the
imaginary time $\tau$, that we can vary. All else are fixed by 
factorization schemes. If the factorization scheme can accurately reproduce
the imaginary time evolution of the wave function, the resulting energy
must fall monotonically from the initial
energy toward the exact ground state energy with increasing $\tau$. 
To the extent that these wave function are not the exact imginary time 
wave function, the energy will eventually rise again. Thus for each
wave function there is a optimal $\tau$ where it will 
be ``closest" to the exact ground state. 
  
To test the quality of the above wave functions 
we use them to describe the ground state of
a strongly correlated quantum system of $N$ $^4$He atoms interacting via
a two-body Aziz HFDHE2 potential.~\cite{aziz79}
%
%
At equilibrium, the system is in a liquid state and has a density of
 $\rho \sigma^3=0.365 \ (\sigma=2.556 \AA)$.
The simplest description of the ground state
is McMillan's Jastrow wave function,
\begin{equation}
\Phi(R)=\exp\left[-\sum_{i>j}^{N} u(r_{ij})\right]  \ \ \ ; \ \ \
u(r)=\frac{1}{2} \left(\frac{b}{r}\right)^5  \ ,
\label{mcmillanwf}
\end{equation}
with $b=1.2\,\sigma$. We will us this wave function as our
initial wave function in all our simulations. 

%
%

For all three wave functions, the expectation 
value of the Hamiltonian can be computed from
\be E=\frac{\int dR \ \Psi(R,\tau) {H} 
\Psi(R,\tau)}{\int dR \ |\Psi(R,\tau)|^2}.
\nonumber
\ee 
The iterated wave functions simply require more
integration variables. For example, in the case of the linear
and quadratic wave function, the above 
can be expressed as
\begin{equation}
E=\int dR \ dS_{L} \ dS_{R} \
    p(R,S_{L},S_{R}) \
    E_L(R,S_{L},S_{R}) \ ,
\label{energy2}
\end{equation}
where   $ p(R,S_{L},S_{R})$ is the probability density function,
$ E_L(R,S_{L},S_{R})$ is the local energy, and
$S_{L,R}$ are the respective left (L), right (R) auxiliary, or shadow,
variables. For the quartic wave function, 
the corresponding expressions for
the probability density function
and energy expectation value are similar but with 
the addition of two more auxiliary shadow variables $S^{\prime}_{L,R}$.

We use the Metropolis Monte Carlo algorithm~\cite{metropolis} to sample 
the probability density from a 9N and 15N-dimensional configuration
space, corresponding to two and four sets of shadow coordinates in the
case of linear/quadratic and quartic wave functions respectively.
In these computations, the Metropolis steps are subdivided in two
parts. First, one attempts to move real particle coordinates at random
inside cubical boxes of side length $\Delta$. Second, analogous 
attempts are made to move shadow coordinates inside cubical boxes of
side length $\Delta_{sh}$.
For instance, in the case of the quadratic wave function, we first attempted
to move all the $R$ coordinates, then move the shadow 
coordinates $\{ S_{L} \}$, then $\{ S_{R} \}$.
The parameters $\Delta$ and $\Delta_{sh}$ were adjusted so that the
acceptance ratio for both particle and shadow moves was nearly $50\%$.

In addition to the ground state variational energy, we have also computed 
the radial distribution function $g(r)$, and its Fourier
transform, the structure factor $S(k)$.
These quantities are spherical averages and have been computed for both
the real particles and the shadow coordinates.
The radial distribution function is defined by
\begin{equation}
g(r)=\frac{1}{N \rho} \sum_{i \neq j}^{N} 
     \left\langle \delta(|{\bf r}_i-{\bf r}_j-{\bf r}|) \right\rangle  \ ,
\label{radial}
\end{equation}
where the angular brackets denote an average with respect to
$|\Psi(R,\tau)|^2$ and $\rho$ is the particle density.
The structure factor $S(k)$ is obtained from the average
$\frac{1}{N} \langle \rho_{-{\bf k}} \rho_{{\bf k}} \rangle$, where
$\rho_{{\bf k}}= \sum_{j=1}^{N} \exp(-i {\bf k}\cdot {\bf r}_j)$, a procedure
which is only possible on a discrete set of ${\bf k}$ values allowed
by the periodic boundary conditions.

All simulations presented in this work have been done with $N=108$
atoms of $^4$He in a cubic box with periodic boundary conditions.
To enforce periodicity all correlations smoothly go to zero at a 
cutoff distance, $r_c=L/2$, equal half the side of the simulation box
according to the replacement
\begin{equation}
f(r) \rightarrow f(r)+f(2 r_c-r)-2 f(r_c)  \ .
\label{smooth}
\end{equation}

%
%

\begin{figure}
\begin{center}
\leavevmode
\hbox{%
\epsfxsize=8cm
\epsffile{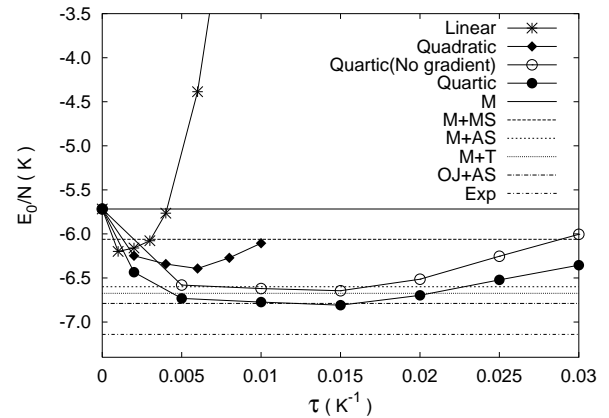}}
\end{center}
\caption{  The Ground state energy per particle in Kelvin for $^4$He at the
           experimental equilibrium density ($\rho \sigma^3=0.365$) using
           the Aziz HFDHE2 potential as a function of the parameter $\tau$.
	   Monte Carlo results from using various short-time evolved wave 
	   functions are as indicated. All simulations have been done for 
	   $N=108$ particles. M indicates a McMillan wave function energy.
           M+MS, M+AS, M+T, OJ+AS refers to various 
           variational Monte Carlo (VMC)
           results  in the literature, see text for details.
           }
\label{fig1}
\end{figure}

In Fig.~\ref{fig1} we show the equilibrium energy per particle 
for liquid $^4$He for various short time evolved wave functions 
as function of the imaginary time parameter $\tau$. Other results from 
literature are also indicated for comparison:
$M+MS$ is the energy obtained by a shadow
wave function having a pure repulsive McMillan (M) 
pseudopotential~\cite{mcmillan} of fifth
power law form for both particles and shadows.~\cite{vitielloprb}
M+AS is the energy obtained by a shadow wave function with an attractive
shadow-shadow pseudopotential of scaled Aziz HFDHE2 
potential (AS) form.~\cite{macfarland}
OJ+AS refers to a shadow wave function with an optimized Jastrow 
particle-particle pseudopotential (OJ) and scaled Aziz HFDHE2 
shadow-shadow pseudopotential (AS).~\cite{macfarland}
GFMC is the Green's Function Monte Carlo calculations
with Mcmillan form for importance and starting
function.~\cite{gfmc}
The experimental value is taken from Roach et al~\cite{roach}.

As expected, each of our factorized wave function reaches
an energy minimum with increasing value of $\tau$. The flatness and
depth of the energy minimun improve markedly with the order of 
the wave function. The linear wave function has a shallow and narrow
minimum at $\tau=0.002$ and only 
improves upon McMillan's result ($\tau=0$) by $\approx 0.3$ K.
The minimum of the quadratic wave function is much better at
$\tau=0.006$ with a value of $-6.393$ K.
The quartic wave function's energy minimum extends further
out to $\tau=0.015$ attaining $-6.809$ K, which is lower than all existing 
variational Monte Carlo calculations that we are aware of. 
To demonstrate the necessity of the gradient term, we have also plotted 
results obtained from (\ref{china}) without the gradient term in 
the potential. In the present case, gradient term is responsible for 
$\approx 50\%$ of the improvement from that of the quadratic wave function.

To give an quantatitive comparison, we summarize various ground state 
equilibrium energies for $^4$He in Table~\ref{tabliquid}. 
  
\begin{figure}
\begin{center}
\leavevmode
\hbox{%
\epsfxsize=8cm
\epsffile{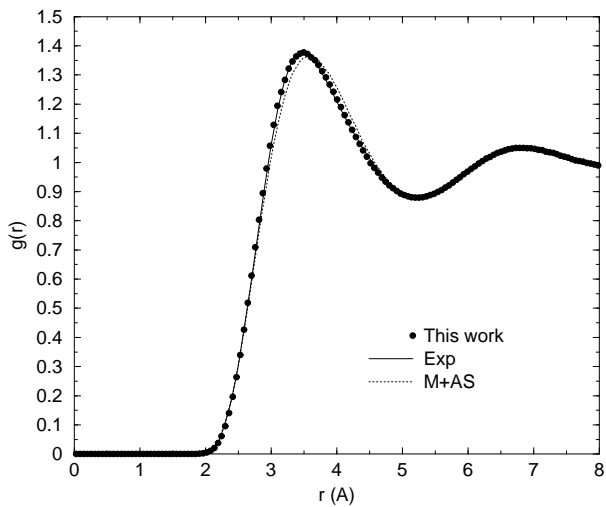}}
\end{center}
\caption[] {The pair distribution function for liquid $^4$He at the
            equilibrium density $\rho \sigma^3=0.365$ after a VMC
            simulations with N=108 particles.
            The filled circles show the $g(r)$ of this work that
            is compared with the respective $g(r)$ obtained from
            the M+AS wave function (dotted line) and the experimental $g(r)$ as
            reported by Svensson et al~\cite{svensson} 
           (solid line)
            obtained at
            saturated vapor pressure at a temperature $T=1.0 \ K$. }
\label{gofr}
\end{figure}

In Fig.~\ref{gofr} we show the equilibrium pair distribution function $g(r)$
for $^4$He as obtained from the quartic wave function.
This $g(r)$ is compared with the respective $g(r)$ obtained from 
the M+AS shadow wave function and the experimental one of 
Svensson et al~\cite{svensson}
obtained by neutron diffraction at saturated vapor pressure at $T=1.0 \ K$.
It is known~\cite{macfarland} that the M+AS curve differs from the
erxperimental one because it predicts a diminished nearest-neighbor
maximum and the entire curve is shifted by about $0.1 \AA$ to larger
values of $r$ compared to the experimental results.
The pair distribution function that we obtained 
is in excellent agreement with the experimental one.

\begin{figure}
\begin{center}
\leavevmode
\hbox{%
\epsfxsize=8cm
\epsffile{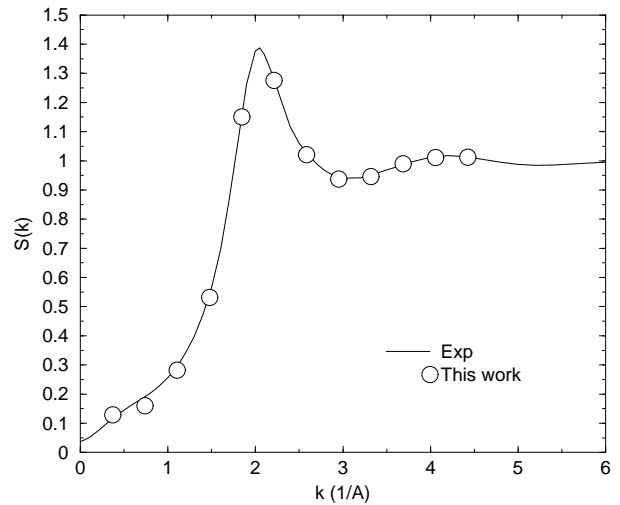}}
\end{center}
\caption[]{Static structure factor $S(k)$ of liquid $^4$He 
           at equilibrium density $\rho \sigma^3=0.365$.
           The filled circles show our results for $S(k)$ obtained
           from the formula
           $S(k)=\frac{1}{N} \langle \rho_{-{\bf k}} \rho_{{\bf k}} \rangle$.
           The solid line denotes the experimental
           results reported by Svensson and co-workers~\cite{svensson}
           obtained at saturated vapor pressure by means of neutron
           diffraction at temperature $T=1.0 \ K$.}
\label{sofk}
\end{figure}

In Fig.~\ref{sofk} we show $S(k)$ at equilibrium density
$\rho \sigma^3=0.365$ as obtained from the
quartic wave function. The experimental $S(k)$ shown in this figure
is the result reported by Svensson et al~\cite{svensson}.
The overall agreement between our short-time evolved structure 
factor with experiment is excellent except at small $k$.
This is not unexpected because our imaginary time is still
rather short for the wave function to develop
the necessary long-range correlation to produce
the linear behavior~\cite{reattoT} of $S(k)$ 
observed in bulk $^4$He.

In this work, based on recent findings on forward time steps 
decomposition schemes, we have implemented a fourth order Short-Time-Evolved
wave function for describing the ground state of strongly interacting 
quantum systems. Our approach is systematic, free of arbitrary parameters,
and can be applied to any general quantum many-body problem.
In the case of liquid $^4$He, we have produced ground state energy 
and structure results better than any existing VMC calculations, 
but without the use of complicated branching processes as in DMC or GFMC. 
Since the anti-symmetric requirement on Fermion
wave functions can be more easily implemented on the variational
level, our quartic wave functions may be of great utility in studying 
Fermi systems. 

Our second order wave function is similar in structure to
the class of {\bf shadow wave functions}~\cite{vitiello}, except that our
wave function follows directly from the second order factorization 
scheme without any particular adjustment of pseudopotential or 
scale functions. Our use of a positive factorization scheme to produce a 
much improved fourth order wave function demonstrates that there is a 
systematic way of improving this class of wave functions by introducting 
more shadow coordinates. Currently, there is no known sixth order 
forward factorization schemes, and hence no sixth order Short-Time-Evolved
wave function is possible.



\begin{table}[t]
\caption[]{ Energies of liquid $^4$He at the experimental equilibrium density
            ($\rho \sigma^3=0.365$ ; $\sigma=2.556 \AA$)
            and at zero temperature. 
            VMC indicates a variational Monte Carlo calculation
            with the indicated wave function.
            All simulations use the Aziz HFDHE2 potential and have been 
            performed for systems of $N=108$ particles.
            The M+MS results are taken from
            Vitiello et al~\cite{vitielloprb}. 
            The M+AS and OJ-AS results are taken from 
            MacFarland et al~\cite{macfarland},
            The GFMC results are taken from Kalos et al~\cite{gfmc}.
            The experimental data are taken from 
            Roach et al~\cite{roach}.
            The energies are given in 
            Kelvin per particle. }
\begin{center}
\begin{tabular}{|c|c|c|} 
\hline  
Method &Trial function                &Energy (K)    \\ \hline   
VMC    &M+MS                          &-6.061 $\pm$ 0.025 
                                                         \\ \hline
VMC    &M+AS                          &-6.599 $\pm$ 0.034  
                                                         \\ \hline
VMC    &OJ+AS                         &-6.789 $\pm$ 0.023
                                                         \\ \hline
VMC    &Linear                        &-6.144 $\pm$ 0.092 
                                                         \\ \hline
VMC    &Quadratic                     &-6.393 $\pm$ 0.021 
                                                         \\ \hline
VMC    &No Grad                       &-6.644 $\pm$ 0.026 
                                                         \\ \hline
VMC    &Quartic                       &-6.809 $\pm$ 0.017 
                                                         \\ \hline
GFMC   &                              &-7.120 $\pm$ 0.024     
                                                         \\ \hline
Experiment &                          &-7.140        
                                                         \\ \hline 
\end{tabular}
\end{center}
\label{tabliquid}
\end{table}



\begin{thebibliography}{999}


\bibitem{rey}P. J. Reynolds, D. M. Ceperley, B. J. Alder, and W. A. Lester,
              J. Chem. Phys. {\bf 77}, 5593 (1982).
\bibitem{mosk}J. W. Moskowitz, K. E. Schmidt, M. E. Lee, and M. H. Kalos,
               J. Chem. Phys. {\bf 77}, 349 (1982).
\bibitem{chindmc} Siu A. Chin, Phys. Rev. {\bf A42}, 6991 (1990).
\bibitem{gfmc} M.H. Kalos, M.A. Lee, P.A. Whitlock, G.V. Chester,
                Phys. Rev. B {\bf 24}, 115 (1981).
\bibitem{forest}E. Forest and R. D. Ruth, Physica D {\bf 43}, 105 (1990).
\bibitem{creutz}M. Creutz and A. Gocksch, Phys. Rev. Letts. {\bf 63}, 9 (1989).
\bibitem{candy}J. Candy and W. Rozmus, J. Comp. Phys. {\bf 92}, 230 (1991).
\bibitem{suzuki}M. Suzuki, Phys. Lett. {\bf A146}, 319 (1990); {\bf 165}, 
                387 (1992).
\bibitem{yoshida}H. Yoshida, Phys. Lett. {\bf A150}, 262 (1990).
\bibitem{manda}R. I. McLaclan and P. Atela, Nonlinearity, {\bf 5}, 542 (1991).
\bibitem{mclachlan}R. I. McLachlan, SIAM J. Sci. Comput. {\bf 16}, 151 (1995).
\bibitem{koseleff}P. V. Koseleff, `Exhaustive search of symplectic integrators 
   using computer algebra', in {\it Integration algorithms and classical 
		  mechanics}, Fields Inst. Commun., 10, Amer. Math. Soc., 
		  Providence, RI, 103 (1996).
\bibitem{sheng}Q. Sheng, IMA Journal of numberical anaysis, {\bf 9}, 
              199 (1989). 
\bibitem{suzukinogo}M. Suzuki, J. Math. Phys. {\bf 32}, 400 (1991).
\bibitem{goldman}D. Goldman and T. J. Kaper, SIAM J. Numer. Anal.,{ \bf 33}, 
                   349 (1996).
\bibitem{suzfour}M. Suzuki, {\it Computer Simulation Studies in 
            Condensed Matter Physics VIII},
           eds, D. Landau, K. Mon and H. Shuttler (Springler, Berlin, 1996).
\bibitem{chin} S.A. Chin, Physics Letters {\bf A} 226, 344 (1997).
\bibitem{kos}P.V. Koseleff, in {\it Applied algebra, algebraic algorithms and 
		error-correcting codes} (San Juan, PR, 1993), Lecture Notes in 
		Comput. Sci., 673, Springer, Berlin, P.213-230 (1993).
\bibitem{chinsym} Siu A. Chin and C. R. Chen, ``Forward Symplectic 
        Integrators for 
        Solving Gravitational Few-Body Problems", arXiv: astro-ph/0304223.
\bibitem{aziz79} R.A. Aziz, V.P.S. Nain, J.S. Carley, W.L. Taylor,
                 G.T. McConville, J. Chem. Phys. {\bf 70}, 4330 (1979).
\bibitem{metropolis} N. Metropolis, A.W. Rosenbluth, M.N. Rosenbluth,
           A.M. Teller, E. Teller, J. Chem. Phys. {\bf 21}, 1087 (1953).
\bibitem{mcmillan} W.L. McMillan, Phys. Rev. {\bf 138A} , 442 (1965).
\bibitem{vitielloprb} S.A. Vitiello, K.J. Runge, G.V. Chester, M.H. Kalos, 
                      Phys. Rev. B {\bf 42}, 228 (1990).  
\bibitem{macfarland} T. MacFarland, S.A. Vitiello, L. Reatto, G.V. Chester,
                     M.H. Kalos, Phys. Rev. B {\bf 50}, 13577 (1994).  
\bibitem{roach} P. T. Roach, J. B. Ketterson, C. W. Woo,
                Phys. Rev. A {\bf 2}, 543 (1970).
\bibitem{svensson} E.C. Svensson, V.F. Sears, A.D.B. Woods, P.Martel,
                   Phys. Rev. B {\bf 21}, 3638 (1980).  
\bibitem{reattoT} L. Reatto, G.V. Chester,
                 Phys. Rev. {\bf 155}, 88 (1967).  
\bibitem{vitiello} S. Vitiello, K. Runge, M.H. Kalos, 
                   Phys. Rev. Lett. {\bf 60}, 1970 (1988).  

%

\end{thebibliography}
\end{document}